\documentclass[12pt,a4paper]{article}
\usepackage{graphicx,cite}
\newcommand{\Frac}[2]{\frac{\displaystyle #1}{\displaystyle #2}}

\newcommand\lsim{\mathrel{\rlap{\lower4pt\hbox{\hskip1pt$\sim$}}
    \raise1pt\hbox{$<$}}}
\newcommand\gsim{\mathrel{\rlap{\lower4pt\hbox{\hskip1pt$\sim$}}
    \raise1pt\hbox{$>$}}}

\newcommand{\bea}{\begin{eqnarray}}
\newcommand{\eea}{\end{eqnarray}}
\newcommand{\beq}{\begin{equation}}
\newcommand{\eeq}{\end{equation}}

\usepackage{amssymb}
\usepackage{graphicx}
\usepackage{epsfig}
\usepackage{latexsym}
\usepackage{amsmath}
\newif\ifpdf
\ifx\pdfoutput\undefined
\pdffalse % we are not running PDFLaTeX
\else
\pdfoutput=1 % we are running PDFLaTeX
\pdftrue
\fi

\newcommand{\fs}{\; .}
\newcommand{\co}{\; ,}  
\newcommand{\non}{\nonumber}
\newcommand{\no}{\nonumber \\}

\newcommand{\nc}{{N_{\!c}}}

\newcommand{\br}{\langle}  %bra
\newcommand{\ke}{\rangle} %ket

\newcommand{\Fz}{F_0} 
\newcommand{\Bz}{B_0}

\newcommand{\uu}{\br  0|\bar u u | 0 \ke_0 }

\newcommand{\Sres}{{\cal S}}
\newcommand{\Pres}{{\cal P}}

\newcommand{\MS}{M_S}
\newcommand{\MP}{M_P}

\newcommand{\MV}{M_V}

\newcommand{\model}{{\Sres\Pres}}
\newcommand{\Lm}{L^\model}
\newcommand{\Cm}{C^\model}

\begin{document}
\begin{titlepage}

%%%%%%%%%%%%%%%%%%%%%%%%%%%%%%%%%%%%%%%%%%%
%Some more stuff to get graphics to work
\ifpdf
\DeclareGraphicsExtensions{.pdf, .jpg}
\else
\DeclareGraphicsExtensions{.eps, .jpg}
\fi

\vspace*{-1.8cm}
\begin{flushright}
{\small\sf  Caltech MAP-306 \\UWThPh-2005-2 \\ IFIC/05-12\\FTUV/05-0309\\}
\end{flushright}

\vspace*{0.7cm} 
\begin{center}
{\Large\bf The $\langle S P P \rangle$ Green Function  and  \\
$   $ \\
SU(3) Breaking in $K_{\ell 3}$ Decays$^*$}
\\[20mm]

{\normalsize\bf \sc V. Cirigliano$^{1}$, G. Ecker$^2$, M.~Eidem\"uller$^{3,4}$,}
\\[10pt]
{\normalsize\bf \sc R. Kaiser$^2$, A. Pich$^{3}$ and J. Portol\'es$^{3}$  }\\

\vspace{1.4cm} 
${}^{1)}$ {\em California Institute of Technology
Pasadena, CA 91125, USA}\\[10pt]
${}^{2)}$ {\em Institut f\"ur Theoretische Physik, University of Vienna,
 A-1090 Vienna, Austria} \\[10pt]
${}^{3)}$ {\em Departament de F\'{\i}sica Te\`orica, IFIC, CSIC --- 
Universitat de Val\`encia \\ 
Edifici d'Instituts de Paterna, Apt. Correus 22085, E-46071 
Val\`encia, Spain} \\[10pt]
${}^{4)}$ {\em Instituto de F\'{\i}sica, Universidade de S\~ao Paulo, 
C.P. 66318, 05315-970 S\~ao Paulo, Brazil} \\
\end{center}

\vfill
\begin{abstract}

Using the $1/N_C$ expansion scheme and truncating the hadronic
spectrum to the lowest-lying resonances, we match a meromorphic
approximation to the $\langle S P P \rangle$ Green function onto QCD
by imposing the correct large-momentum falloff, both off-shell and on
the relevant hadron mass shells. In this way we determine a number of
chiral low-energy constants of $O(p^6)$, in particular the ones
governing $SU(3)$ breaking in the $K_{\ell 3}$ vector form factor at zero 
momentum transfer.
The main result of our matching procedure is that the known loop
contributions largely dominate the corrections of $O(p^6)$ to
$f_{+} (0)$. We discuss the implications of our final value 
$f_{+}^{K^0 \pi^-} (0)=0.984 \pm 0.012$ for the
extraction of $V_{us}$ from $K_{\ell 3}$ decays.
\end{abstract}

\vfill
\noindent 
*~Work supported in part by HPRN-CT2002-00311 (EURIDICE) and by
Acciones Integradas, HU2002-0044 (MCYT, Spain), Project No. 19/2003 
(Austria).
\end{titlepage}

\section{Introduction}

Chiral perturbation theory~\cite{Weinberg:1978kz,Gasser:1983yg,
Gasser:1984gg} (CHPT) 
is the effective theory describing the low-energy expansion of QCD Green
functions. It is a fundamental tool in low-energy hadron
phenomenology.  State-of-the-art calculations involve expansions to
NNLO ($p^6$) in external momenta and quark
masses~\cite{Bijnens:2004pk}.  Given the large number of low-energy
constants (LECs) appearing in the CHPT effective Lagrangian to order
$p^6$~\cite{Fearing:1994ga,Bijnens:1999sh,Bijnens:1999hw},
in order to
retain predictive power it is highly desirable to develop a
non-perturbative framework to match the effective low-energy
description to QCD and to estimate the LECs.
In this work we use the $1/N_C$ expansion framework together with
the Minimal Hadronic Ansatz~\cite{Peris:1998nj,Pich:2002xy,deRafael:2002tj} 
to study the matching
between low and high energies for the $\langle S P P \rangle$
three-point function of one scalar and two pseudoscalar densities in
the chiral limit.

The purpose of our investigation is twofold.  First, we wish to extend
the $1/N_C$-motivated matching scheme that has proved successful in
other cases~\cite{Peris:1998nj,Pich:2002xy,deRafael:2002tj,Moussallam:1994at,
Moussallam:1997xx,Knecht:2001xc,Ruiz-Femenia:2003hm,Bijnens:2003rc,
Cirigliano:2004ue} to
the $\langle S P P \rangle$ three-point function.  The scheme entails
the construction of a hadronic interpolation between the known low- and
high-momentum regimes, dictated respectively by chiral symmetry and by
the QCD asymptotic behaviour.  The approximations involve the
choice of the hadronic content and the corresponding set of 
short-distance constraints to be satisfied.
We truncate the spectrum to the lowest-lying resonance multiplet per
channel, based on the observation that the low-lying hadronic spectrum
has the largest impact on the LECs.  This choice defines
our hadronic ansatz as the most general meromorphic function with
poles corresponding to Goldstone bosons and $\Sres$, $\Pres$
resonances. 
Concerning the short-distance behaviour, the leading power in the OPE 
for the $\langle S P P \rangle$ Green function displays
anomalous scaling, making the explicit matching
with a meromorphic function problematic. In our analysis we therefore 
impose that the behaviour of our ansatz at large momenta is not worse than
the one required by QCD. These constraints are then fully compatible
with the asymptotic vanishing of those hadronic form
factors that we need to consider for fixing the relevant LECs.

Beyond the general aspects mentioned above, the $\langle S P P \rangle$
Green function is of special phenomenological interest, as its
low-energy behaviour allows to determine the LECs governing SU(3) breaking
in $K_{\ell 3}$ decays to $O(p^6)$. In turn, $K_{\ell 3}$ decays
offer the possibility of a precise determination of the CKM
mixing parameter $V_{us}$, and thus an accurate test of CKM
unitarity when combined with knowledge of
$V_{ud}$~\cite{Battaglia:2003in}. Our aim is to explore the
quantitative implications of our matching framework for $K_{\ell 3}$
decays and to assess the attendant uncertainty.

The material in this paper is organized as follows. In
Sec.~\ref{sec:SPP} we describe in detail the matching procedure for
the $\langle SPP \rangle$ correlator. We give its low- and
high-momentum limits, the interpolating form and we discuss the
implications for the LECs.  In Sec.~\ref{sec:kl3} we then review
the status of CHPT calculations of $K_{\ell 3}$ form factors and the
impact of our findings on the local contribution of $O(p^6)$.  Finally, in
Sec.~\ref{sec:conclusions} we summarize our main results and
conclusions.  

\section{The $\langle SPP \rangle$ Green function}
\label{sec:SPP}
Our starting point is the Fourier transform of the octet $\langle SPP
\rangle$ Green function in massless QCD. On account of SU(3) and
$C$ invariance it is given by a single scalar function\footnote{ The
quark currents are normalized according to
$$
P^a(x)  =  \bar{q}(x) i \gamma_5 \lambda^a q (x) \; \co\;\;  
S^a(x) =  \bar{q}(x) \lambda^a q (x) \co
$$
with $ \br \lambda^a \lambda^b \ke =2 \delta^{ab} $; furthermore,
 $d^{abc} = \frac{1}{4} \br \lambda^a
\{\lambda^b, \lambda^c \} \ke$.  },
\begin{equation}\label{SPP-def}
i^2\! \int \! dx\,  dy \, e^{i px +i qy + irz} \br 0 | T S^a (x) P^b
(y) P^c (z) | 0 \ke =  d^{abc} \,\Pi_{SPP} (p^2,q^2,r^2) \co
\end{equation}
with $p+q+r = 0$. Bose symmetry implies that the function is symmetric
in its second and third arguments
\begin{equation}\label{SPP-bose}
\Pi_{SPP} (s,u,t) = \Pi_{SPP} (s,t,u) \fs
\end{equation}

%%%%%%%%%%%%%%%%%%%%%%%%%%%%%%%%%%%%%%%%%%%%%%%%
%%%%%%%%%%%%%%%%%%%%%%%%%%%%%%%%%%%%%%%%%%%%%%%%
\subsection{Chiral symmetry}

The low-energy expansion of the function $\Pi_{SPP}$ may be worked out
in CHPT: 
\begin{eqnarray}\label{SPP-chiral}
\Pi_{SPP}(s,t,u) & = & \frac{(2 \Bz)^3}{t u} \Big\{  \Fz^2  +4 L_5 \,
s + 4 (4L_8-L_5) \, (t+u) \\
& & -8 C_{12} \, s^2 + 8 (2 C_{12}+C_{34}+C_{38}) \, s (t+u)   -
8 (C_{12}+C_{34}-C_{38}) \,  (t^2+u^2) \Big\}  \no 
&&  - (4\Bz)^3(2C_{12}-4C_{31}-2C_{34}+2C_{38}-C_{94})+  O(p^2) \, ,\non
\end{eqnarray}
where the loop contributions have been discarded. The leading-order 
contribution to $\Pi_{SPP}$ depends exclusively on the pion
decay constant $\Fz$ and on the quark condensate, $ \Bz = - \uu
/\Fz^2 $ in the chiral limit $m_u =m_d
=m_s =0$. The coefficients of the higher-order contributions are
written in terms of the coupling constants $L_i$ and $C_i$ 
\cite{Gasser:1984gg,Bijnens:1999sh}\footnote{In the literature, there
  exist different conventions as to the normalization of the $C_i$  
\cite{Bijnens:1999sh,Bijnens:1999hw}. We prefer to
work with the $C_i$ of mass dimension $-2$ since they have the
canonical large-$\nc$ behaviour.}.

%%%%%%%%%%%%%%%%%%%%%%%%%%%%%%%%%%%%%%%%%%%%%%%%
%%%%%%%%%%%%%%%%%%%%%%%%%%%%%%%%%%%%%%%%%%%%%%%%
\subsection{Asymptotic behaviour}

The Operator Product Expansion (OPE) implies the following short-distance
behaviour of the $\langle SPP \rangle$ Green function  when 
$s,t,u \to \infty$:
\begin{equation}\label{SPP-OPE}
\Pi_{SPP} (s,t,u) \rightarrow  2 \uu \,  \frac{s^2 -(t-u)^2}{s t  u} \co 
\end{equation}
to leading order in inverse powers of momenta, but to zeroth order in the
strong coupling constant. In fact, the anomalous dimensions of the
(pseudo)scalar currents and of the quark condensate imply nontrivial
Wilson coefficients in (\ref{SPP-OPE}) that ensure the correct scale
dependence and modify the asymptotic behaviour by logarithms. We come
back to this important point in subsection \ref{subsec:sd}.

In the case where only two of the three coordinates $x$, $y$, $z$ 
approach each other, the analysis reduces to the OPE
of pairs of quark currents and yields in momentum
space\footnote{ The term given explicitly does in fact involve the 
two-point function $\br 0 |T A^a_\mu(x) P^b(y) |0 \ke$. However, 
for vanishing quark
masses, that correlation function coincides with its short-distance
limit,
$$
\br 0 |T A^a_\mu(x) P^b(y) |0  \ke = - 2 i \delta^{ab} 
\uu \partial_\mu \Delta_0(x-y) +O(m_q)   \ . 
$$
}   
\begin{align}
\label{sd:p fix} \Pi_{SPP} (p^2,q^2,(p+q)^2) &= O(q^{-2})   \co  
	& p \; {\mbox{fixed}}  \co \; q \to \infty   \co \\
\label{sd:q fix} \Pi_{SPP} (p^2,q^2,(p+q)^2) &= -  8 \uu  \, \frac{p
  q}{p^2 q^2} + O(p^{-2}) \co 	& q \; {\mbox{fixed}}  \co \; p \to \infty   \fs 
\end{align} 

There is one additional constraint that follows from chiral symmetry:
chiral Ward identities and pion pole dominance lead to the relation 
\begin{equation}\label{SS-PP}
\lim_{t \to 0 }\, \frac{t}{2\Bz}   \, \Pi_{SPP}(s,t,s) =  
\Pi_{SS}(s)-  \Pi_{PP}(s)  \, ,
\end{equation}
where we have introduced the scalar two-point functions $\Pi_{SS}$ and
$ \Pi_{PP}$ according to
\begin{eqnarray}
i \! \int \! dx\,  e^{i p(x-y)} \br 0 | T S^a (x) S^b (y) | 0 \ke  & =
&   
\delta^{ab} \,\Pi_{SS} (p^2) \co \\
i \! \int \! dx\, e^{i p(x-y)} \br 0 | T P^a (x) P^b (y)  | 0 \ke  & =
&  
\delta^{ab} \,\Pi_{PP} (p^2) \fs \non
\end{eqnarray}
It is well known that the OPE for $T S^a (x)
S^b (y)$ coincides with the one for $T P^a (x) P^b (y)$ in the chiral
limit up to terms of order $(x-y)^{-2}$. We thus deduce
\begin{equation}\label{sd:SS-PP}
\lim_{t \to 0 }\, \frac{t}{2\Bz}   \, \Pi_{SPP}(s,t,s) = O(s^{-2})  \co
\end{equation}
when $s \to \infty$, not conflicting with (\ref{sd:q fix}).

A different class of asymptotic constraints comes from
considering the behaviour of various hadronic form factors at high
momentum transfer.  Form factors may be obtained from $\Pi_{SPP}$ by
extracting the residues of the appropriate double poles.  As an explicit
example, the scalar form factor of the pion $F_S^{\pi\pi} (s) $ is
given by the residue of the double pole at $t = u = 0$,
\begin{equation}\label{FpiS_1}
F_S^{\pi\pi} (s) = \lim_{t , u \to 0} \frac{tu }{(2\Bz\Fz)^2} \, 
\Pi_{SPP} (s,t,u) \fs
\end{equation}
The asymptotic condition~\cite{Lepage:1979zb} in this case reads
\begin{equation}\label{FpiS_2}
\lim_{s \to \infty}  F_S^{\pi\pi}(s)  = 0  \ . 
\end{equation}
Similar constraints exist for other (transition) form factors.

%%%%%%%%%%%%%%%%%%%%%%%%%%%%%%%%%%%%%%%%%%%%%%%
%%%%%%%%%%%%%%%%%%%%%%%%%%%%%%%%%%%%%%%%%%%%%%%
\subsection{Model approximation} 

In the large-$N_C$ limit the only singularities in $\Pi_{SPP}$ are
single-meson poles. In this spirit, and truncating the spectrum to one 
resonance per channel, we construct a meromorphic
approximation to the function $\Pi_{SPP}$ with explicit scalar
($\Sres$)  and pseudoscalar ($\Pres$) poles (besides the pion),
\begin{equation}\label{SPP-model}
\Pi_{SPP}^\model(s,t,u)  = 8 \Bz^3 \Fz^2 \MS^2 \MP^4 
\frac{P_0+P_1+P_2+P_3+P_4}{[\MS^2 -s][-t][-u][\MP^2-t][\MP^2-u]} \co
 \end{equation}
where the $P_n$ are polynomials of degree $n$ in $s$, $t$, $u$:
\begin{equation}
 P_n = \sum_{k=0}^{n} \sum_{l=0}^{k} c_{n-k,k-l,l} \, s^{n-k} t^{k-l}
 u^l  \fs
\end{equation}

The ansatz in Eq.~(\ref{SPP-model}) represents the most
general expression for the given particle content (fixing the
denominator) that does not violate the short-distance behaviour
$\Pi_{SPP} = O(p^{-2})$ required by the OPE
(\ref{SPP-OPE}). The normalization is chosen such that $P_0
\equiv c_{000}=1$ yields the correct low-energy limit. Since the
function $\Pi_{SPP}$ is symmetric under the interchange of $t$ and $u$
we have in addition
\begin{equation}
 c_{kml}  =  c_{klm} \fs
\end{equation}
We are thus left with 21 parameters $c_{100} , \ldots , c_{022}$ to be
determined.

It is straightforward to work out the implications of our model for
the chiral coupling constants that enter in Eq.~(\ref{SPP-chiral}).
Aside from $c_{000}=1$, the expressions for the LECs
involve the coefficients in $P_1$ and, in case of the $C_i$, also
those from $P_2$:
\begin{eqnarray}\label{Lecs-model}
\Lm_5 & = & \frac{\Fz^2}{4}\left[\frac{1}{\MS^2} +c_{100} \right] \; \co \;\; 
\Lm_8  =  \frac{\Fz^2}{16}\left[\frac{1}{\MS^2} +\frac{1}{\MP^2} +
c_{100} +c_{010} \right] \co\\
\Cm_{12} & = & - \frac{1}{2 \MS^2} \Lm_5 - \frac{\Fz^2}{8} c_{200} \co \no
\Cm_{34} & = &  \frac{1}{2}\left[\frac{1}{\MS^2} + \frac{1}{\MP^2}
  \right]    
 \Lm_5 +\left[\frac{1}{\MS^2}- \frac{1}{\MP^2} \right]  \Lm_8 - 
\frac{\Fz^2}{16} \left[  \frac{1}{\MS^2\MP^2} -3 c_{200}-c_{110}+ 
c_{020} \right] \co \no
\Cm_{38}  & = &  \left[\frac{1}{\MS^2}+\frac{1}{\MP^2} \right]  \Lm_8 - 
\frac{\Fz^2}{16}  \left[ \frac{1}{\MS^2\MP^2} -
  c_{200}-c_{110}-c_{020} \right]   \fs \non
\end{eqnarray}
The coupling constants $C_{31}$ and $ C_{94} $ are not fixed
individually. For the combination $C_{31}+ \frac{1}{4} C_{94}$ one
finds
\begin{equation}
\Cm_{31}+ \frac{1}{4} \Cm_{94}  =  - \Cm_{34}+ \Cm_{38}-
\frac{\Fz^2}{32}  
\left[ \frac{1}{\MP^4}  + 2c_{020} -c_{011} \right] \fs
\end{equation} 

Before moving on to the asymptotic constraints, let us observe that the
relation between $\Pi_{SPP}$ and $\Pi_{SS}-\Pi_{PP}$ given in
Eq.~(\ref{SS-PP}) implies that the combination $c_{100}+c_{010}$
relevant for $L_8$ is given in terms of quantities that specify the
two-point functions. 
In fact, defining $c_m$ and $d_m$ in terms of the one-particle matrix
elements of the scalar and pseudoscalar currents 
\cite{Ecker:1988te},
\begin{equation}
\label{eq:cmcd}
|\br 0 | S^a| \Sres^b \ke|   =  \delta^{ab}\,4\sqrt{2}  \Bz\,  
c_m   \; \co \;\; 
|\br 0 | P^a | \Pres^b \ke|  =    \delta^{ab}\,4\sqrt{2}  \Bz\,  d_m  \ , 
\end{equation} 
one derives from Eq.~(\ref{SS-PP}) 
\begin{equation}\label{SS-PP:cklm1}
\frac{1}{\MS^2} +\frac{1}{\MP^2} +c_{100} +c_{010} = 
\frac{8 }{\Fz^2} \left[ \frac{c_m^2}{\MS^2}- 
 \frac{d_m^2}{ \MP^2}  \right]  \ , 
\end{equation}
which implies the  well-known result~\cite{Ecker:1988te} 
$\Lm_8  = 1/2 \, ( c_m^2/\MS^2 - d_m^2/\MP^2 )$. 

The $\langle S P P \rangle$ Green function has also been studied in 
Ref.~\cite{Bijnens:2003rc} within a ladder resummation inspired
hadronic model. Phenomenological consequences such as the LECs of
$O(p^6)$ were not considered in that approach.

%%%%%%%%%%%%%%%%%%%%%%%%%%%%%%%%%%%%%%%%%%%%%%%%
%%%%%%%%%%%%%%%%%%%%%%%%%%%%%%%%%%%%%%%%%%%%%%%%
\subsection{Implementing asymptotic constraints}
\label{subsec:sd}

The aim of the present investigation is to obtain information on the
chiral coupling constants by exposing the ansatz in
Eq.~(\ref{SPP-model}) to suitable constraints implied by QCD
asymptotic behaviour. In particular, it has been proven successful
\cite{Peris:1998nj,Pich:2002xy,deRafael:2002tj,Moussallam:1994at,
Moussallam:1997xx,Knecht:2001xc, Ruiz-Femenia:2003hm,
 Bijnens:2003rc,Cirigliano:2004ue} to match a
meromorphic representation to the QCD short-distance behaviour, given
in this case by Eqs.~(\ref{SPP-OPE}), (\ref{sd:p fix}), and (\ref{sd:q
fix}). Here we encounter a problem related to the missing Wilson
coefficients in those relations. For instance, matching to
Eq.~(\ref{SPP-OPE}) would imply that the nine
coefficients in $P_4$ are fixed completely,
\begin{equation}\label{cklm-OPE-1}
2 c_{211} = c_{022} = -2 c_{031} = \frac{1}{2 \Bz^2 \MS^2 \MP^4}  \co
\end{equation}
with all other coefficients in $P_4$ being zero.

The problem alluded to is manifest in Eq.~(\ref{cklm-OPE-1}), which would
imply that non-vanishing coefficients $c_{klm}$ depend on the running
scale of QCD in the same manner as the (quark condensate)$^{-2}$. 
Including the appropriate Wilson coefficients in Eqs.~(\ref{SPP-OPE})
and (\ref{sd:q fix}) would repair the scale dependence but it would
make the $c_{klm}$ momentum dependent at the same time invalidating
our ansatz. Moreover, that momentum dependence would involve the
strong coupling in the non-perturbative regime. It is also obvious
that the logarithmic dependence induced by
nontrivial Wilson coefficients can never be matched by a meromorphic
approximation with a finite number of resonances (see, e.g., 
Ref.~\cite{Shifman:2000jv}).

Therefore, we have to find an alternative set of criteria to determine 
the parameters of our model ansatz and the relevant 
LECs. Of course, this discussion is not limited to 
$\langle SPP \rangle$ but applies to all Green functions with
anomalous scaling to leading power in inverse momenta.

1. While matching onto the short-distance result leads
to the problems discussed above, we certainly wish to ensure that the
short-distance behaviour of our ansatz is not worse than what is
predicted in QCD. The behaviour $\Pi_{SPP} = O(p^{-2})$ (up to logarithms) 
was built into our ansatz from the start. There are, however,
nontrivial restrictions that follow from the limits in which the
momenta are treated asymmetrically:
\begin{align}
\Pi_{SPP} (p^2,q^2,(p+q)^2) & = O(q^{-2})   \co  &  p \;
   {\mbox{fixed}}  
\co \; q \to \infty \co  \tag{\ref{sd:p fix}'} \\
\Pi_{SPP} (p^2,q^2,(p+q)^2) & =  O(p^{-1})  \co  &  q \;
   {\mbox{fixed}}  
\co \; p \to \infty  \fs \tag{\ref{sd:q fix}'} 
\end{align} 
Our model (\ref{SPP-model}) reproduces this behaviour, provided the
following relations hold:
\begin{eqnarray}
\label{eq:HOPE}
c_{400}+c_{310}+c_{220}+c_{130}+c_{040} & = & 0 \co \\  	% (x.1) SS-PP  qNWT
8c_{400}+3c_{310}-c_{130} & = & 0 \co \no
% (x.2)               qNWT
6c_{400}+3c_{310}+c_{220}& = & 0 \co \no
% (x.3)               qNWT
c_{310}+c_{211}+c_{121}+c_{031} & = & 0 \co \no
% (x.4)               qNWT
2c_{040}+2c_{031}+c_{022} & = & 0 \co \no
% (x.5)               pNWT
c_{300}+c_{210}+c_{120}+c_{030}& = & 0 \fs \non
% (x.6) SS-PP  qNWT
\end{eqnarray}

2. More conditions arise from the short-distance behaviour of $\Pi_{SS}(s)-
\Pi_{PP}(s)$ in (\ref{sd:SS-PP}).  First, the parameters $c_m$ and
$d_m$ must satisfy \cite{Moussallam:1994at,Golterman:1999au}
\begin{equation}
\label{eq:swsr}
c_m^2 -  d_m^2 = \frac{\Fz^2}{8}  \co 
\end{equation}
a relation analogous to the first Weinberg sum rule
\cite{Weinberg:1967kj}. In addition, Eq.~(\ref{sd:SS-PP}) requires
that the following combinations of coefficients vanish:
\begin{align}\label{SS-PP:cklm2}
c_{200}+c_{110} +c_{020}  &= 0  \co  \\
c_{300}+c_{210}+c_{120}+c_{030} &= 0  \co  \no 
c_{400}+c_{310}+c_{220}+c_{130}+c_{040} &= 0  \fs \non
\end{align}

3. The asymptotic vanishing of the scalar form factor of
the pion requires
\begin{equation}\label{FpiS:cklm}
c_{100}=c_{200}=c_{300}=c_{400}=0  \ . 
\end{equation}

When combining  the restrictions from the pion scalar form factor in
Eq.~(\ref{FpiS:cklm}), the scalar/ pseudoscalar Weinberg sum rule in
Eq.~(\ref{eq:swsr}), the OPE conditions in Eq.~(\ref{eq:HOPE}), and 
Eq.~(\ref{SS-PP:cklm1}), we are left with a total of 9 undetermined
parameters: 
\begin{align}
P_2  & =c_{110}[s(t+u)-t^2-u^2] + c_{011}tu \co \\
P_3  &= c_{210}[s^2(t+u)-t^3-u^3]+c_{111}stu +c_{021}[t+u]tu+
c_{120}[s(t^2+u^2)-t^3-u^3] \co \no
P_4 &= c_{310}[s^3(t+u)-3s^2(t^2+u^2)+3s(t^3+u^3)-(t^2+3tu+u^2)(t-u)^2]\no
&\;\;\;\;+ c_{211}[s^2-(t-u)^2]tu+c_{121}[s(t+u)-(t-u)^2]tu  \fs \non 
\end{align}

4. Once the conditions from one-pion transition form factors $ \langle \pi | S
| \Pres \rangle $ ($c_{310}=c_{210}=0$, $c_{110} = - M_P^2 c_{120}$)
and $ \langle \pi | P | \Sres \rangle$ ($c_{120}=0$) are included, the
ansatz reduces further to
\begin{eqnarray}
P_2 & = & c_{011}tu \co \label{eq:poly}
\\
P_3 & = & [c_{111}s+c_{021}(t+u)]tu \co \no
P_4 & = & [c_{211}(s^2-(t-u)^2)+c_{121}(s(t+u)-(t-u)^2)]tu \ .  \non 
\end{eqnarray}
The constraints discussed so far are sufficient to fix 
the LECs of order $p^4$ and $p^6$ within our scheme.

We could fix additional parameters in the polynomials
  (\ref{eq:poly}) by also considering the various form factors of the
  $\Sres$ and  $\Pres$ resonance states. As pointed out in
  Ref.~\cite{Bijnens:2003rc}, this procedure in general leads to a
  representation in conflict with the OPE constraints. We do not dwell
  on this issue further since it is of no concern for our purpose of 
  determining the LECs occurring up to $O(p^6)$.

%%%%%%%%%%%%%%%%%%%%%%%%%%%%%%%%%%%%%%%%%%%%%%%%
%%%%%%%%%%%%%%%%%%%%%%%%%%%%%%%%%%%%%%%%%%%%%%%%

\subsection{Low-energy constants}
\label{sec:lecs}

The constraints enumerated above determine the LECs $L_5, L_8, C_{12},
C_{34}$ and $C_{38}$ in terms of resonance masses and couplings.  
\begin{itemize}
\item The asymptotic vanishing of the pion scalar form factor fixes
$L_5$ and $C_{12}$ through $c_{100} = c_{200}=0$. The scalar form factor
defined in Eq.~(\ref{FpiS_1}) takes the simple form
\begin{equation}\label{FpiS_3}
F_S^{\pi\pi}(s) = F_S^{\pi\pi}(0) \frac{\MS^2}{\MS^2-s} \fs
\end{equation}
The coupling constants describing the dependence on the variable $s$
in the low-energy expansion of
$\Pi_{SPP}$, viz. $L_5$ and $C_{12}$ in Eq.~(\ref{SPP-chiral}), are
determined by the momentum dependence of the scalar form factor
alone. Within the single-scalar resonance approximation that implies 
(see also \cite{Leutwyler:1989pn})
\begin{equation}
L _5 = \frac{\Fz^2}{4 \MS^2} \; \co \;\; C_{12} =  - \frac{\Fz^2}{8 \MS^4}  \fs 
\end{equation}       
In the vector sector the analogous consideration leads to the
well-known predictions
\cite{Ecker:1989yg,Moussallam:1997xx,Knecht:2001xc,Cirigliano:2004ue}
\begin{equation}
L _9 = \frac{\Fz^2}{2 \MV^2} \; \co \;\; C_{88} -C_{90} =  - \frac{\Fz^2}{4 \MV^4}  \fs 
\end{equation}   

\item Enforcing the correct short-distance behaviour of
$\Pi_{SS}(s)- \Pi_{PP}(s)$ by virtue of the first of
Eqs.~(\ref{SS-PP:cklm2}) determines the value of $C_{38}$.
\item The asymptotic vanishing of one-pion transition form factors 
$ \langle \pi | S | \Pres \rangle $ and $ \langle \pi | P | \Sres
  \rangle$ implies $c_{110} = c_{020} =0$ and thus determines $C_{34}$.
\end{itemize}
The combination $C_{31} + 1/4 \, C_{94}$ remains undetermined, as 
the coefficient $c_{011}$ is not fixed by the constraints we 
have considered.

Within our framework the relevant LECs of $O(p^6)$ are determined in
terms of the scalar and pseudoscalar octet masses ($M_S$ and $M_P$),
the pion decay constant $\Fz$  and the couplings $c_m$ and $d_m$
defined in Eq.~(\ref{eq:cmcd}). Using also the Weinberg-like sum
rule (\ref{eq:swsr}), we obtain
\begin{eqnarray}\label{Lecs-model-2}
\Lm_5 & = & \frac{\Fz^2}{4 \, \MS^2}  \; \co \;\; 
\Lm_8  =  \frac{1}{2}\left(\frac{c_m^2}{\MS^2} - \frac{d_m^2}{\MP^2} \right)
\co  \no 
\Cm_{12} & = & - \frac{\Fz^2}{8 \MS^4}   
\co \no
\Cm_{34} & = &  \frac{3 \, \Fz^2}{16 \MS^4} + \frac{d_m^2}{2} 
\left(\frac{1}{\MS^2} - \frac{1}{\MP^2} \right)^2  
\co \no
\Cm_{38}  & = &  
\frac{\Fz^2}{16 \MS^4} + \frac{d_m^2}{2} 
\left(\frac{1}{\MS^4} - \frac{1}{\MP^4} \right)  
\fs 
\end{eqnarray}
To estimate these couplings, we need
numerical values for the input parameters.  For the pion decay
constant we use the physical value $\Fz = F_\pi = 92.4$ MeV.  
The coupling $d_m$ can be fixed by studying the pion scalar form factor
away from the chiral limit~\cite{Pich:2002xy}, resulting in $d_m =
\Fz/(2 \sqrt{2})$ (or $c_m = \Fz/2$), which we take as central value.
As for the mass parameters, spectroscopy and chiral
symmetry~\cite{pdg04,Cirigliano:2003yq} suggest a central value 
$M_P= 1.3$ GeV, while $M_S$ is more controversial.  The analysis of
Ref.~\cite{Cirigliano:2003yq} would suggest $M_S =1.48$ GeV for the 
lightest scalar nonet that survives in the large-$N_C$ limit.  This 
result is supported by recent lattice calculations (see for example 
Ref.~\cite{Prelovsek:2004jp} and references therein). It
implies a value of $\Lm_{5} \simeq 10^{-3}$, in good
agreement with most recent fits to the $O(p^4)$ 
LECs~\cite{Amoros:2001cp}.

\begin{table}[ht]
\begin{center}
\begin{tabular}{|c|c|c|c|}
\hline
 &  & & \\[-.2cm] 
 & $C_i \cdot 10^{4}$ GeV$^{2}$ & $C_i \cdot (4\pi)^{4} F_\pi^{2}$ 
 & $\delta C_i \cdot 10^{4}$ GeV$^{2}$ \\[.2cm] 
\hline 
 &  &  & \\[-.2cm] 
\hspace*{.5cm}  $\Cm_{12}$ \hspace*{.5cm}  &  \hspace*{.5cm} 
$- 4.4 $ \hspace*{.5cm}  & \hspace*{.5cm} $-0.09$
\hspace*{.5cm} & $1.6$ \\[.2cm] 
$\Cm_{34}$ &  $ 6.6 $ & $0.14$ & $4.7$ \\[.2cm] 
$\Cm_{38}$ &  $ 2.5 $ & $0.05$ & $2.2$ \\[.2cm]
\hline
\end{tabular}
\end{center}
\caption{Numerical values for the LECs of $O(p^6)$ in GeV$^{-2}$ and
  in natural units $(4\pi)^{-4} F_\pi^{-2}$ for $M_S=1.25$ GeV. The 
  last column contains the variations of the $C_i(\mu)$ for $M_\eta 
  \le \mu \le 1$ GeV. More precisely, we display the quantities
  $\delta C_i= {\rm  max}  \left\{ |C_{i}(M_\rho) - C_{i} (M_\eta)|, 
  |C_{i}(M_\rho) - C_{i} (1 \, {\rm GeV})| \right\} $, using the
  $L_i^r(\mu)$ from fit 10 in Ref.~\cite{Amoros:2001cp}.}
\label{tab:Ci}
\end{table}
In the estimates below, we use $M_P = 1.3$ GeV and we vary
$M_S$ between 1 and 1.5 GeV.  
The numerical values for $M_S=1.25$ GeV are collected in Table~\ref{tab:Ci}.
The above results represent the leading term in the large-$N_C$
expansion of the couplings (within our simplified scheme of truncating
the spectrum to the lowest-lying resonances).  One way to estimate the
size of subleading corrections in $1/N_C$ is to look at the
renormalization scale dependence of the couplings.  This
effect is formally higher order in $1/N_C$ and a leading-order
estimate is unable to provide the scale at which the expressions
(\ref{Lecs-model-2}) apply.  At $O(p^4)$, resonance saturation
works well for $\mu = M_\rho$. On this basis, a crude estimate
of the uncertainty is given by the variation of $C_i(\mu)$ for 
$\mu$ between $M_\eta$ and 1 GeV.

In this way, we obtain the uncertainties shown in Table~\ref{tab:Ci}. 
Two comments are in order here. First of all, it is
not the uncertainty of any given LEC that matters\footnote{After all, 
the choice of LECs is basis dependent.} but the overall
scale dependence of a measurable quantity, as will be discussed
below for the $K_{\ell 3}$ vector form factor at $t=0$. Secondly,
the values in Table~\ref{tab:Ci} are of course very sensitive to
the scalar resonance mass, while the scale dependence stays 
the same. This strong 
dependence on $M_S$ seems rather discomforting at first sight but, 
as above, physical observables may exhibit a smoother dependence 
as we will demonstrate in the following section.

\section{SU(3) breaking in $K_{\ell 3}$ decays and $V_{us}$}
\label{sec:kl3}

We now investigate the consequences of our results for the estimate of
SU(3) breaking in $K_{\ell 3}$ form factors.  As is well known,
$K_{\ell 3}$ decays offer one of the most accurate determinations of
the CKM element $V_{us}$. After the recent re-evaluation of
radiative corrections~\cite{radcorr1,radcorr2} and the new experimental
results \cite{Alexopoulos:2004sw,Alexopoulos:2004sy,Franzini:2004kb,Lai:2004bt} 
(see Ref.~\cite{Mescia:2004xd} for a review of the present
experimental and theoretical status), the main uncertainty in
extracting $V_{us}$ comes from theoretical calculations of the vector 
form factor $f_{+}^{K^0 \pi^-}(0)$ at zero momentum transfer defined by
\begin{equation} \label{eq:fff}
\langle \pi^- (p_\pi) | \bar{s} \gamma_\mu u | K^0 (p_K) \rangle = 
f_{+}^{K^0 \pi^-} (t)  \, (p_K + p_\pi)_\mu  + 
f_{-}^{K^0 \pi^-} (t)  \, (p_K - p_\pi)_\mu  \ , 
\end{equation}
where $t = (p_K - p_\pi)^2$.  Here we are interested in the $SU(3)$
breaking corrections to $f_{+}^{K^0 \pi^-}(0)$.
We break up the form factor in terms of its expansion in quark masses:
\begin{equation}
f_{+}^{K^0 \pi^-} (0) = 1 + f_{p^4} + f_{p^6} + \dots \fs 
\end{equation}
Deviations from unity (the octet symmetry limit) are of second order
in $SU(3)$ breaking~\cite{agth}.  The first correction arises to
$O(p^4)$ in CHPT: a finite one-loop
contribution~\cite{Gasser:1984ux,Leutwyler:1984je} determines
$f_{p^4}= - 0.0227$ in terms of $F_\pi$, $M_K$ and $M_\pi$, with
essentially no uncertainty.  The $p^6$ term receives contributions
from pure two-loop diagrams, one-loop diagrams with insertion of one
vertex from the $p^4$ effective Lagrangian, and pure tree-level
diagrams with two insertions from the $p^4$ Lagrangian or one
insertion from the $p^6$ Lagrangian~\cite{Post:2001si,Bijnens:2003uy}:
\begin{equation}
f_{p^6} = f_{p^6}^{\rm 2-loops} (\mu) +   
f_{p^6}^{L_i \times {\rm loop}} (\mu) 
+  f_{p^6}^{\rm tree} (\mu)    \ . 
\end{equation}
Individual components depend on the chiral renormalization scale
$\mu$, their sum being scale independent.  Formally speaking, the
three contributions scale as $O(1/N_C^2)$, $O(1/N_C)$,
and $O(1)$, respectively.  Although our main concern here is with 
$f_{p^6}^{\rm tree}$, the other terms have to be accounted for in a consistent
phenomenological analysis, as infrared logs tend to upset the $1/N_C$ counting.
Using as reference scale $\mu=M_\rho=0.77$ GeV and the $L_i$ from fit
10 in Ref.~\cite{Amoros:2001cp}, one has~\cite{Bijnens:2003uy}: 
\begin{eqnarray}
f_{p^6}^{\rm 2-loops} (M_\rho) &=&  0.0113  \; ,  \\   
f_{p^6}^{L_i \times {\rm loop}} (M_\rho) &=& - 0.0020 \pm 0.0005 \ .
\label{eq:Liloop}     
\end{eqnarray}
Note that we have subtracted the tree-level piece proportional to 
$L_5 \times L_5$ from the corresponding quantity $\Delta(0)$ in
Ref.~\cite{Bijnens:2003uy}. 

\begin{figure}[!t]
\centering

\begin{picture}(300,175)  
\put(110,65){\makebox(50,50){\epsfig{figure=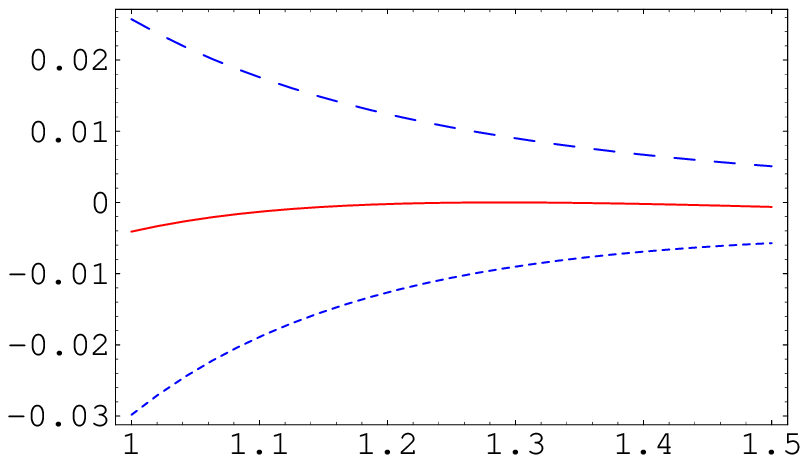,width=10cm}}}
\put(260,-5){
{\large 
$M_S \ ({\rm GeV})$ 
}
}
\put(-55,160){
{\large 
$ f_{p^6}^{\rm tree} (M_\rho)  $ 
}
}
\put(110,140){
{\small
$L_5 \times L_5 / F_\pi^2 $
}
}
\put(100,50){
{\small
$- (C_{12} + C_{34})$
}
}
\end{picture}
\caption{
\label{fig:ftree}
We display $f_{p^6}^{\rm tree} (M_\rho)$ according to
Eq.~(\ref{eq:tree2}) as a function of $M_S$ for $M_P=1.3$ GeV 
(solid line). 
We also plot the two components according to 
Eq.~(\ref{eq:tree1}): the dashed line represents  the 
term proportional to $L_5 \times L_5$, while the 
dotted line represents the term proportional to 
$- (C_{12} + C_{34})$.
}
\end{figure}

The explicit form for the tree-level 
contribution is then~\cite{Bijnens:1998yu,Bijnens:2003uy}
\begin{equation}
\label{eq:tree1}
f_{p^6}^{\rm tree} (M_\rho) = 
8 \frac{\left( M_K^2 - M_\pi^2 \right)^2}{F_\pi^2}  
\, \left[\frac{\left(L_5^r (M_\rho) \right)^2}{F_\pi^2} - 
C_{12}^r (M_\rho) - C_{34}^r (M_\rho) \right] . 
\end{equation}
Upon substituting $L_5^r (M_\rho) \to \Lm_5 $ and $C_{12,34}^r (M_\rho)
\to \Cm_{12,34}$ one gets
\begin{equation}
\label{eq:tree2}
f_{p^6}^{\rm tree} (M_\rho) = 
%- 4 \frac{d_m^2}{F_\pi^2} \, \frac{\left( M_K^2 - M_\pi^2 \right)^2}{M_S^4}  
%\, \left( 1 - \frac{M_S^2}{M_P^2} \right)^2  = 
-  \frac{\left( M_K^2 - M_\pi^2 \right)^2}{2 \, M_S^4}  
\, \left( 1 - \frac{M_S^2}{M_P^2} \right)^2   \ , 
\end{equation}
where we used $d_m = F_\pi/(2 \sqrt{2})$.  
In Fig.~\ref{fig:ftree} we plot as a function of $M_S$ our full result
for $f_{p^6}^{\rm tree} (M_\rho)$ (solid line) as well as its two
components: the $L_5 \times L_5$ term (dashed line) and the $C_{12} +
C_{34}$ piece (dotted line).  The two contributions tend to largely
cancel each other, reducing the full result to $\sim 10 \%$ of each
individual term.  As a consequence, the ambiguity related to
$M_S$ is strongly reduced, given the size of the resulting effect.
This is simply a consequence of treating all tree-level
contributions to $f_{p^6}$ on the same footing, as suggested by the
$1/N_C$ counting.  

%%%%%%%%%%%%%%%%%%%%%%%%%%%%%%%%%%%%%%%%%%%%%%%%%%%
%%%%%%%%%%%%%%%%%%%%%%%%%%%%%%%%%%%%%%%%%%%%%%%%%%%%%%%%%

In addition to varying $M_S$ in the range 1 GeV $\le M_S \le $ 1.5
GeV, we need to estimate the intrinsic uncertainty due to our
choice of $\mu=M_\rho$ as matching point for $f_{p^6}^{\rm tree}
(M_\rho)$.  We do this by performing the matching for any $\mu \in
[M_\eta, 1 \, {\rm GeV}]$ and then running the result back to
$\mu=M_\rho$ via the renormalization group~\cite{Bijnens:1999hw}.
This way we find $ \delta f_{p^6}^{\rm tree} (1/N_C) = \pm 0.008 $.
Accounting also for the uncertainty in $M_S$, we get
\begin{eqnarray} \label{eq:fp6n}
f_{p^6}^{\rm tree} (M_\rho) &=& - 0.002  \pm 0.008_{\, 1/N_C} \pm 
0.002_{\, M_S}  \fs \label{eq:fp6} 
\end{eqnarray}
We now discuss the two main features of our result: 
\begin{itemize} 
\item[i.] The smallness of the local contribution 
$f_{p^6}^{\rm tree} (M_\rho)$. 
\item [ii.] The size of formally subleading terms in the
$1/N_C$ expansion: the scale dependence of the LECs and the loop 
contribution to $f_{p^6}$.
\end{itemize}  

The naive expectation for the size of the local contribution to
$f_{p^6}$ is $(M_K^2 - M_\pi^2)^2/M_S^4 \sim 5 \cdot 10^{-2}\, 
{\rm GeV}^4/M_S^4$. An order of magnitude is lost through
the factor $1/2 (1 - M_S^2/M_P^2)^2 < 0.1$ (for resonance masses in
the range considered) in Eq.~(\ref{eq:tree2}). This extra suppression
is a consequence of imposing, within the particle content of our
ansatz, the correct asymptotic behaviour for the two- and one-pion
form factors $\langle \pi | S | \pi \rangle$, $\langle \pi | S | \Pres 
\rangle$ and $\langle \pi | P | \Sres \rangle$. 

We do not have a
complete answer to the question whether this suppression persists when
using a more sophisticated ansatz. However, we have examined the
stability of our result in two different directions. Omitting
the pseudoscalar resonances altogether in our ansatz (\ref{SPP-model})
gives rise to a solution that is equivalent to setting $d_m=0$ in 
Eqs.~(\ref{Lecs-model-2}). There is a complete destructive interference in
this case: the scalar contributions cancel in
Eq.~(\ref{eq:tree1}) implying $f_{p^6}^{\rm tree} (M_\rho)=0$ instead of 
Eq.~(\ref{eq:tree2}). On the other hand, adding an additional
pseudoscalar nonet leads to a straightforward generalization 
of Eq.~(\ref{eq:tree2}). Instead of
\begin{equation} 
d_m^2  \left( 1 - \frac{M_S^2}{M_P^2} \right)^2  \qquad {\rm with}
  \qquad d_m^2=c_m^2 - \displaystyle\frac{F_\pi^2}{8}=
  \displaystyle\frac{F_\pi^2}{8} \; , 
\end{equation}
one gets by including a second pseudoscalar multiplet with mass
$M_{P^\prime}$ and coupling $d_m^{\prime}$
\begin{equation} 
d_m^2  \left( 1 - \frac{M_S^2}{M_P^2} \right)^2 + 
d_m^{\prime \,2}  \left( 1 - \frac{M_S^2}{M_{P^\prime}^2} \right)^2
\qquad {\rm with} \qquad d_m^2 + d_m^{\prime 2} =
  \displaystyle\frac{F_\pi^2}{8} \fs  
\end{equation} 
Assuming $d_m^{\prime \,2} \le d_m^2$, $M_S=1.25$ GeV and taking 
$M_{P^\prime}$ in the
range 1.5 $\to$ 2 GeV, $|f_{p^6}^{\rm tree} (M_\rho)|$ may increase by
up to 0.002. Thus, additional pseudoscalar multiplets do not
modify the general result of a small tree-level contribution
$f_{p^6}^{\rm tree} (M_\rho)$. 
\par
The second issue concerns higher-order corrections in the $1/N_C$
expansion.  
We have taken the variation of the LECs with the renormalization scale
as a measure of those corrections. The relatively big variation for
$\mu$ in the range $M_\eta \le \mu \le 1$ GeV is due to large
infrared logs that typically occur in the scalar sector. 
This also explains why loop contributions cannot be 
neglected in this case.  
\par
Adding all uncertainties linearly, but omitting the small
error of the loop contribution (\ref{eq:Liloop}), we get
finally\footnote{Estimates of the uncertainty due to higher-order 
corrections beyond $O(p^6)$ \cite{Bijnens:2003uy} essentially do
not modify the final result (\ref{eq:ff0}) for $f_{+}^{K^0 \pi^-} (0)$
when adding the errors in quadrature.}
\begin{eqnarray}
f_{p^6}^{\rm tree} (M_\rho) &=& - 0.002  \pm 0.008_{\, 1/N_C} \pm 
0.002_{\, M_S} \,\mbox{}_{- 0.002}^{+0.000} \,\mbox{}_{\, P^\prime} \ ,   \\
f_{p^6} &=& 0.007 \pm 0.012 \ , \label{eq:fp6final} \\  
f_{+}^{K^0 \pi^-} (0) &=&  0.984 \pm 0.012 \fs
\label{eq:ff0} 
\end{eqnarray}
Our final result (\ref{eq:ff0}) differs from other 
determinations~\cite{Leutwyler:1984je,Jamin:2004re,Becirevic:2004ya}.
This difference is due to the small central value for $f_{p^6}^{\rm
tree}$, which appears to be a generic consequence of a few-resonance
approximation. When combined with the size and sign of the
loop contribution of $O(p^6)$~\cite{Bijnens:2003uy}, our central value 
for $f_{p^6}$ in (\ref{eq:fp6final}) is positive in contrast to most other
estimates, e.g., $f_{p^6}^{\rm LR} = - 0.016 \pm 0.008$ of Leutwyler
and Roos \cite{Leutwyler:1984je}. 
\par
For the purpose of illustration, we use the recent $K_L$ branching
ratio measurements of KTeV~\cite{Alexopoulos:2004sw}, together
with their results on $K_{e3}$ form factors
where the curvature in $f_{+}^{K^0 \pi^-} (t)$ has been included in
the analysis~\cite{Alexopoulos:2004sy}. Using also the
recent precise determination of the $K_L$ lifetime
\cite{Franzini:2004kb}, one finds \cite{HNHonnef} 
$ \ f_{+}^{K^0 \pi^-} (0) \cdot |V_{us}| = 0.2166 \pm 0.0010$.  Our
result for $ f_{+}^{K^0 \pi^-} (0) $ in (\ref{eq:ff0}) then implies
\begin{equation}
\label{eq:vus}
|V_{us}| = 0.2201 \pm 0.0027_{f_+(0)} \pm 0.0010_{\rm exp} \fs 
\end{equation}
The central value is smaller than the one from CKM
unitarity, using the most recent value of $V_{ud}$
\cite{Czarnecki:2004cw}. It is also
smaller than what one would obtain using the same experimental input and
$f_{+}^{K^0 \pi^-} (0)$ from
Refs.~\cite{Leutwyler:1984je,Jamin:2004re,Becirevic:2004ya}.  On the
other hand, our result is in better agreement with alternative
extractions of $V_{us}$ from $\tau$ decays~\cite{Gamiz:2004ar} and
$K_{\ell 2}/\pi_{\ell 2}$~\cite{Marciano:2004uf}.
\par
The LECs obtained in this article determine also the deviation from the 
original Callan-Treiman relation 
\cite{Callan:1966hu} 
\begin{equation}
\Delta_{CT} \, = \, f_0^{K^0 \pi^-}
(M_K^2-M_{\pi}^2) \, - \, \Frac{F_K}{F_{\pi}} \; ,
\end{equation}
involving the scalar form factor $f_0(t) = f_+(t) + 
t f_-(t)/(M_K^2-M_{\pi}^2)$.
The tree-level contribution of $O(p^6)$ 
is given by \cite{Jamin:2004re}
\begin{eqnarray}
\Delta_{CT}^{\rm{tree},p^6} & = & 16 \, \Frac{M_{\pi}^2}{F_{\pi}^2} \, 
( M_K^2 - M_{\pi}^2) \, ( 2 \, C_{12}^r \, + \, C_{34}^r ) 
\, = \,  \Frac{M_{\pi}^2 (M_K^2 - M_{\pi}^2)}{M_P^4} \, \left( \,
1 \, - \,  2 \,  \Frac{M_P^2}{M_S^2} \,  \right) \; .
\end{eqnarray}
With the values of $M_S$ and $M_P$ considered before,
$\Delta_{CT}^{\rm{tree},p^6}$ is negative and small (a few times
$10^{-3}$ in magnitude).
Finally, these LECs also provide a prediction for the slope\footnote{The
slope $\lambda_0$ is defined by
$$
f_0(t) = f_+(0) \left[ \, 1 \, + \, \lambda_0 \, \Frac{t}{M_{\pi^+}^2} + \ldots
\right] \; .
$$} of the scalar form factor. In the representation for $\lambda_0$
given in Ref.~\cite{Bijnens:2003uy}, the LECs appear again in the
combination $2 \,C_{12}  +  C_{34}$.
Including the loop contributions \cite{Bijnens:2003uy}, we obtain  
\begin{equation}
\lambda_0 = 0.013 \pm 0.002_{\, 1/N_C} \pm 
0.001_{\, M_S} = (13 \pm 3)\cdot 10^{-3}\; ,
\end{equation}
in agreement with $\lambda_0 = (17 \pm 4)\cdot 10^{-3}$ from 
Ref.~\cite{Gasser:1984ux}, $\lambda_0 = (15.7 \pm 1.0)\cdot 10^{-3}$ from 
Ref.~\cite{Jamin:2004re}
 and with the value measured by 
KTeV \cite{Alexopoulos:2004sy} in $K^L_{\mu3}$ decays, 
$\lambda_0 = (13.72 \pm 1.31) \cdot 10^{-3}$.

%%%%%%%%%%%%%%%%%%%%%%%%%%%%%%%%%%%%%%%%%%%%%%%%
%%%%%%%%%%%%%%%%%%%%%%%%%%%%%%%%%%%%%%%%%%%%%%%%

\section{Conclusions}
\label{sec:conclusions}

We have constructed a meromorphic approximation to the
$\langle S P P \rangle$ Green function with pole singularities
corresponding to Goldstone modes and lowest-lying scalar ($\Sres$) and
pseudoscalar ($\Pres$) resonances.  The highlights of our analysis are:
\begin{itemize}
\item 
We have shown how the polynomial terms in our ansatz can be fixed by
imposing the correct large-momentum behaviour implied by QCD both
off-shell (OPE constraints) and on the relevant hadron mass shells
(form factor constraints). Because of nontrivial Wilson coefficients,
one cannot match the OPE constraints exactly with a finite number of
resonance poles. We have instead required a large-momentum behaviour 
for our ansatz that is not worse than predicted by QCD. 
For the particle content used, no inconsistencies arise with this set 
of constraints.
\item
This matching procedure has allowed us to determine three of the 
LECs of $O(p^6)$ ($C_{12}$, $C_{34}$, $C_{38}$) appearing in the low-energy 
expansion of $\langle S P P \rangle$ in terms of resonance masses 
$M_{S,P}$ and $F_\pi$. In particular, $C_{12}$ is uniquely determined 
by requiring the correct behaviour of the pion scalar form factor 
$ \langle \pi | S | \pi \rangle $, while $C_{34}$ is fixed by the 
correct scaling of the one-pion form factors 
$ \langle \pi | S | \Pres \rangle $ and $\langle \pi | P | \Sres \rangle$.
\item
We have estimated the uncertainty of the large-$N_C$ matching procedure
by varying the chiral renormalization scale at which the matching is
performed.  While being a subleading effect in the $1/N_C$ counting,
the scale ambiguity $M_\eta \le \mu \le 1$ GeV gives rise to sizable 
uncertainties. For the same reason, loop contributions must be included
in phenomenological applications.
\item
We have explored the impact of our results on the estimate of the
local $p^6$ contribution to $SU(3)$ breaking in $K_{\ell 3}$ decays.
We find that the resulting effect is much smaller than the ratio of
mass scales $(M_K^2 - M_\pi^2)^2/M_S^4 $ would suggest, due to
interfering contributions. When combined with known loop corrections
\cite{Bijnens:2003uy}, our estimate leads to $f_{p^6} = 0.007 
\pm 0.012$, the mean value being opposite in sign 
to most other existing calculations.
Using this input and the most recent experimental results for the 
$K_{e3}$ branching ratio~\cite{Alexopoulos:2004sw} and lifetime for 
$K_L$~\cite{Franzini:2004kb}, we find
$|V_{us}| = 0.2201 \pm 0.0027_{f_+(0)} \pm 0.0010_{\rm exp}$.  
The mean value of our result is smaller than the value inferred from
CKM unitarity $|V_{us}|^{\rm unit.} = 0.2265 \pm
0.0022$, using the most recent determination of $V_{ud}$ 
\cite{Czarnecki:2004cw}. If the recent precision measurement of the
neutron lifetime \cite{Serebrov:2004zf} were confirmed our preferred 
value for $V_{us}$ would however be in perfect agreement with CKM 
unitarity.
\item
We have considered variations of the hadronic ansatz to investigate
the stability of our results. The smallness of the tree-level part
compared to the loop contribution of $O(p^6)$ for $f_{+}^{K^0 \pi^-}
(0)$ appears as generic feature of a few-resonance approximation
for the set of large-momentum constraints considered. In view of the 
significant implications for the determination of $V_{us}$, the
validity of our approach will be further investigated also for other
Green functions. 
\item
Finally, we have also used our results to estimate
the deviation from the Callan--Treiman relation and to calculate the 
slope of the scalar $K_{\ell 3}$ form factor.
\end{itemize}
\vspace*{0.2cm}

\noindent{\large\bf Acknowledgements}
\vspace*{0.3cm} \\
V.C., A.P. and J.P. thank Maarten Golterman and Santi Peris for hospitality
at the Benasque Workshop {\em ``Matching Light Quarks to Hadrons''},
where an early part of this work was carried out. We also 
thank Hans Bijnens, J\"urg Gasser and Helmut Neufeld for useful
information and discussions. V.C. is supported by
a Sherman Fairchild Fellowship from Caltech. M.E. thanks the FAPESP
(Brazil) for financial support. This work has been supported in part by
MCYT (Spain) under Grant FPA2004-00996, by Generalitat Valenciana
(Grants GRUPOS03/013 and GV04B-594) and by ERDF funds from the 
European Commission.

%%%%%%%%%%%%%%%%%%%%%%%%%%%%%%%%%%%%%%%%%%%%%%%%%%%%%%%
%%%%%%%%%%%%%%%%%%%%%%%%%%%%%%%%%%%%%%%%%%%%%%%%%%%


\begin{thebibliography}{10}

\bibitem{Weinberg:1978kz}
%\cite{Weinberg:1978kz}
% \bibitem{Weinberg:1978kz}
S.~Weinberg,
%``Phenomenological Lagrangians,''
Physica A {\bf 96} (1979) 327.
%%CITATION = PHYSA,A96,327;%%

\bibitem{Gasser:1983yg}
%\cite{Gasser:1983yg}
%\bibitem{Gasser:1983yg}
J.~Gasser and H.~Leutwyler,
%``Chiral Perturbation Theory To One Loop,''
Ann. Phys.\  {\bf 158} (1984) 142.
%%CITATION = APNYA,158,142;%%

\bibitem{Gasser:1984gg}
%\cite{Gasser:1984gg}
J.~Gasser and H.~Leutwyler,
%``Chiral Perturbation Theory: Expansions In The Mass Of The Strange Quark,''
Nucl.\ Phys.\ B {\bf 250} (1985) 465.
%%CITATION = NUPHA,B250,465;%%

%\cite{Bijnens:2004pk}
\bibitem{Bijnens:2004pk}
For a recent review see J.~Bijnens,
{\it Chiral meson physics at two loops},
arXiv:hep-ph/0409068.
%%CITATION = HEP-PH 0409068;%%


%\cite{Fearing:1994ga}
\bibitem{Fearing:1994ga}
H.~W.~Fearing and S.~Scherer,
%``Extension of the chiral perturbation theory meson Lagrangian to order
%p(6),''
Phys.\ Rev.\ D {\bf 53} (1996) 315
[arXiv:hep-ph/9408346].
%%CITATION = HEP-PH 9408346;%%

%\cite{Bijnens:1999sh}
\bibitem{Bijnens:1999sh}
J.~Bijnens, G.~Colangelo and G.~Ecker,
%``The mesonic chiral Lagrangian of order p**6,''
JHEP {\bf 9902} (1999) 020
[arXiv:hep-ph/9902437].
%%CITATION = HEP-PH 9902437;%%

%\cite{Bijnens:1999hw}
\bibitem{Bijnens:1999hw}
J.~Bijnens, G.~Colangelo and G.~Ecker,
%``Renormalization of chiral perturbation theory to order p**6,''
Ann. Phys.\  {\bf 280} (2000) 100
[arXiv:hep-ph/9907333].
%%CITATION = HEP-PH 9907333;%%

%\cite{Peris:1998nj}
\bibitem{Peris:1998nj}
S.~Peris, M.~Perrottet and E.~de Rafael,
%``Matching long and short distances in large-N(c) {QCD},''
JHEP {\bf 9805} (1998) 011
[arXiv:hep-ph/9805442];\\
%%CITATION = HEP-PH 9805442;%%
%\cite{Knecht:1999gb}
%\bibitem{Knecht:1999gb}
  M.~Knecht, S.~Peris, M.~Perrottet and E.~de Rafael,
  %``Decay of pseudoscalars into lepton pairs and large N(c) QCD,''
  Phys.\ Rev.\ Lett.\  {\bf 83} (1999) 5230
  [arXiv:hep-ph/9908283];\\
  %%CITATION = HEP-PH 9908283;%%
  %\cite{Peris:2000tw}
%\bibitem{Peris:2000tw}
  S.~Peris, B.~Phily and E.~de Rafael,
  %``Tests of large-N(c) QCD from hadronic tau decay,''
  Phys.\ Rev.\ Lett.\  {\bf 86} (2001) 14
  [arXiv:hep-ph/0007338].
  %%CITATION = HEP-PH 0007338;%%

%\cite{Pich:2002xy}
\bibitem{Pich:2002xy}
%For reviews  see 
A.~Pich, in {\it  Phenomenology of large-$N_C$ QCD}, ed. R.F.~Lebed, 
World Scientific, Singapore 2002, p. 239,
arXiv:hep-ph/0205030.
%%CITATION = HEP-PH 0205030;%%

%\cite{deRafael:2002tj}
\bibitem{deRafael:2002tj}
  E.~de Rafael,
  %``Analytic approaches to kaon physics,''
  Nucl.\ Phys.\ Proc.\ Suppl.\  {\bf 119} (2003) 71
  [arXiv:hep-ph/0210317].
  %%CITATION = HEP-PH 0210317;%%

%\cite{Moussallam:1994at}
\bibitem{Moussallam:1994at}
B.~Moussallam and J.~Stern, in {\it Two-photon physics: From
  DA$\Phi$NE to LEP 200 and beyond}, F. Kapusta and J. Parisi eds., 
World Scientific, Singapore 1994, arXiv:hep-ph/9404353.
%%CITATION = HEP-PH 9404353;%%

%\cite{Moussallam:1997xx}
\bibitem{Moussallam:1997xx}
B.~Moussallam,
%``A sum rule approach to the violation of Dashen's theorem,''
Nucl.\ Phys.\ B {\bf 504} (1997) 381
[arXiv:hep-ph/9701400].
%%CITATION = HEP-PH 9701400;%%

%\cite{Knecht:2001xc}
\bibitem{Knecht:2001xc}
M.~Knecht and A.~Nyffeler,
%``Resonance estimates of O(p**6) low-energy constants and QCD  short-distance
%constraints,''
Eur.\ Phys.\ J.\ C {\bf 21} (2001) 659
[arXiv:hep-ph/0106034].
%%CITATION = HEP-PH 0106034;%%

%\cite{Bijnens:2003rc}
\bibitem{Bijnens:2003rc}
J.~Bijnens, E.~Gamiz, E.~Lipartia and J.~Prades,
%``QCD short-distance constraints and hadronic approximations,''
JHEP {\bf 0304} (2003) 055
[arXiv:hep-ph/0304222].
%%CITATION = HEP-PH 0304222;%%

\bibitem{Ruiz-Femenia:2003hm}
P.~D.~Ruiz-Femenia, A.~Pich and J.~Portol\'es,
%``Odd-intrinsic-parity processes within the resonance effective theory of
%QCD,''
JHEP {\bf 0307} (2003) 003
[arXiv:hep-ph/0306157].
%%CITATION = HEP-PH 0306157;%%

%\cite{Cirigliano:2004ue}
\bibitem{Cirigliano:2004ue}
V.~Cirigliano, G.~Ecker, M.~Eidem\"uller, A.~Pich and J.~Portol\'es,
%``The  Green function in the resonance region,''
Phys.\ Lett.\ B {\bf 596} (2004) 96
[arXiv:hep-ph/0404004].
%%CITATION = HEP-PH 0404004;%%

%\cite{Battaglia:2003in}
\bibitem{Battaglia:2003in}
M.~Battaglia et al., {\it The CKM matrix and the unitarity triangle}, 
arXiv:hep-ph/0304132.
%%CITATION = HEP-PH 0304132;%%

\bibitem{Lepage:1979zb} 
G.P.~Lepage and S.J.~Brodsky, Phys. Lett. B {\bf 87} (1979) 359,
%\bibitem{Lepage:1980fj} 
%G.P.~Lepage and S.J.~Brodsky, 
Phys. Rev. D {\bf 22} (1980) 2157;\\
%\bibitem{Brodsky:1981rp} 
S.J.~Brodsky and G.P.~Lepage, Phys. Rev. D
{\bf 24} (1981) 1808.

%\cite{Ecker:1988te}
\bibitem{Ecker:1988te}
G.~Ecker, J.~Gasser, A.~Pich and E.~de Rafael,
%``The Role Of Resonances In Chiral Perturbation Theory,''
Nucl.\ Phys.\ B {\bf 321} (1989) 311.
%%CITATION = NUPHA,B321,311;%%

\bibitem{Shifman:2000jv}
M.A.~Shifman, in {\it At the frontier of particle physics: 
Handbook of QCD. Boris Ioffe festschrift}, M.~A.~Shifman ed., 
World Scientific, Singapore 2000,
%"Quark-hadron duality",%
arXiv:hep-ph/0009131;\\
%%CITATION = HEP-PH 0009131;%%
%{Golterman:2001pj}
M.~Golterman, S.~Peris, B.~Phily and E.~de Rafael,
%"Testing an approximation to large-N(c) QCD with a toy model",%
JHEP {\bf 01} (2002) 024 [arXiv:hep-ph/0112042].
%%CITATION = HEP-PH 0112042;%%



%\cite{Golterman:1999au}
\bibitem{Golterman:1999au}
M.~F.~L.~Golterman and S.~Peris,
%``The 7/11 rule: An estimate of m(rho)/f(pi),''
Phys.\ Rev.\ D {\bf 61} (2000) 034018
[arXiv:hep-ph/9908252].
%%CITATION = HEP-PH 9908252;%%


%\cite{Weinberg:1967kj}
\bibitem{Weinberg:1967kj}
S.~Weinberg,
%``Precise Relations Between The Spectra Of Vector And Axial Vector Mesons,''
Phys.\ Rev.\ Lett.\  {\bf 18} (1967) 507.
%%CITATION = PRLTA,18,507;%%
%Resonannces


%\cite{Leutwyler:1989pn}
\bibitem{Leutwyler:1989pn}
H.~Leutwyler,
%``How About M (U) = 0?,''
Nucl.\ Phys.\ B {\bf 337} (1990) 108.
%%CITATION = NUPHA,B337,108;%%

%\cite{Ecker:1989yg}
\bibitem{Ecker:1989yg}
G.~Ecker, J.~Gasser, H.~Leutwyler, A.~Pich and E.~de Rafael,
%``Chiral Lagrangians For Massive Spin 1 Fields,''
Phys.\ Lett.\ B {\bf 223} (1989) 425.
%%CITATION = PHLTA,B223,425;%%

\bibitem{pdg04} 
%\cite{Eidelman:2004wy}
%\bibitem{Eidelman:2004wy}
S.~Eidelman et al.  [Particle Data Group Collaboration],
%``Review of particle physics,''
Phys.\ Lett.\ B {\bf 592} (2004) 1.
%%CITATION = PHLTA,B592,1;%%

%\cite{Cirigliano:2003yq}
\bibitem{Cirigliano:2003yq}
V.~Cirigliano, G.~Ecker, H.~Neufeld and A.~Pich,
%``Meson resonances, large N(c) and chiral symmetry,''
JHEP {\bf 0306} (2003) 012
[arXiv:hep-ph/0305311].
%%CITATION = HEP-PH 0305311;%%

%\cite{Prelovsek:2004jp}
\bibitem{Prelovsek:2004jp}
S.~Prelovsek, C.~Dawson, T.~Izubuchi, K.~Orginos and A.~Soni,
%``Scalar meson in dynamical and partially quenched two-flavour QCD: Lattice
%results and chiral loops,''
Phys.\ Rev.\ D {\bf 70} (2004)  094503
[arXiv:hep-lat/0407037].
%%CITATION = HEP-LAT 0407037;%%

%\cite{Amoros:2001cp}
\bibitem{Amoros:2001cp}
G.~Amor\'os, J.~Bijnens and P.~Talavera,
%``QCD isospin breaking in meson masses, decay constants and quark mass
%ratios,''
Nucl.\ Phys.\ B {\bf 602} (2001) 87 [arXiv:hep-ph/0101127].
%%CITATION = HEP-PH 0101127;%%

\bibitem{radcorr1}
%\cite{Cirigliano:2004pv}
%\bibitem{Cirigliano:2004pv}
V.~Cirigliano, H.~Neufeld and H.~Pichl,
%``K(e3) decays and CKM unitarity,''
Eur.\ Phys.\ J.\ C {\bf 35} (2004) 53 [arXiv:hep-ph/0401173]; \\
%%CITATION = HEP-PH 0401173;%%
%\cite{Cirigliano:2001mk}
%\bibitem{Cirigliano:2001mk}
V.~Cirigliano, M.~Knecht, H.~Neufeld, H.~Rupertsberger and P.~Talavera,
%``Radiative corrections to K(l3) decays,''
Eur.\ Phys.\ J.\ C {\bf 23} (2002) 121
[arXiv:hep-ph/0110153].

\bibitem{radcorr2}
%%CITATION = HEP-PH 0110153;%%
%\cite{Andre:2004tk}
%\bibitem{Andre:2004tk}
T.~C.~Andre, {\it Radiative corrections to $K^0_{l3}$ decays},
arXiv:hep-ph/0406006 ; \\
%%CITATION = HEP-PH 0406006;%%
%\cite{Bytev:2002nx}
%\bibitem{Bytev:2002nx}
V.~Bytev, E.~Kuraev, A.~Baratt and J.~Thompson,
%``Radiative corrections to the K+-(e3) decay revised,''
Eur.\ Phys.\ J.\ C {\bf 27} (2003) 57
[Erratum-ibid.\ C {\bf 34} (2004) 523]
[arXiv:hep-ph/0210049].
%%CITATION = HEP-PH 0210049;%%

%\cite{Alexopoulos:2004sw}
\bibitem{Alexopoulos:2004sw}
T.~Alexopoulos et al.  [KTeV Collaboration],
%``A determination of the CKM parameter $|$V(us)$|$,''
Phys.\ Rev.\ Lett.\  {\bf 93} (2004) 181802
[arXiv:hep-ex/0406001].
%%CITATION = HEP-EX 0406001;%%

%\cite{Alexopoulos:2004sy}
\bibitem{Alexopoulos:2004sy}
T.~Alexopoulos {\it et al.}  [KTeV Collaboration],
%``Measurements of semileptonic K(L) decay form factors,''
Phys.\ Rev.\ D {\bf 70} (2004) 092007
[arXiv:hep-ex/0406003].
%%CITATION = HEP-EX 0406003;%%

%\cite{Franzini:2004kb}
\bibitem{Franzini:2004kb}
P.~Franzini, {\it Kaon decays and $V_{us}$},
arXiv:hep-ex/0408150.
%%CITATION = HEP-EX 0408150;%%

%\cite{Lai:2004bt}
\bibitem{Lai:2004bt}
A.~Lai {\it et al.}  [NA48 Collaboration],
%``Measurement of the branching ratio of the decay K(L) $\to$ pi+- e-+ nu and
%extraction of the CKM parameter $|$V(us)$|$,''
Phys.\ Lett.\ B {\bf 602} (2004) 41
[arXiv:hep-ex/0410059].
%%CITATION = HEP-EX 0410059;%%


%\cite{Mescia:2004xd}
\bibitem{Mescia:2004xd}
F.~Mescia, {\it $V_{us}$ from $K_{l3}$ decays},
arXiv:hep-ph/0411097.
%%CITATION = HEP-PH 0411097;%%

\bibitem{agth}
%\cite{Ademollo:1964sr}
%\bibitem{Ademollo:1964sr}
M.~Ademollo and R.~Gatto,
%``Nonrenormalization Theorem For The Strangeness Violating Vector Currents,''
Phys.\ Rev.\ Lett.\  {\bf 13} (1964) 264;\\
%%CITATION = PRLTA,13,264;%%
%\cite{Behrends:1960nf}
% \bibitem{Behrends:1960nf}
R.~E.~Behrends and A.~Sirlin,
%``Effect Of Mass Splittings On The Conserved Vector Current,''
Phys.\ Rev.\ Lett.\  {\bf 4} (1960) 186.
%%CITATION = PRLTA,4,186;%%

%\cite{Gasser:1984ux}
\bibitem{Gasser:1984ux}
J.~Gasser and H.~Leutwyler,
%``Low-Energy Expansion Of Meson Form-Factors,''
Nucl.\ Phys.\ B {\bf 250} (1985) 517.
%%CITATION = NUPHA,B250,517;%%

%\cite{Gasser:1984ux}
%\cite{Leutwyler:1984je}
\bibitem{Leutwyler:1984je}
H.~Leutwyler and M.~Roos,
%``Determination Of The Elements V (Us) And V (Ud) Of The Kobayashi-Maskawa
%Matrix,''
Z.\ Phys.\ C {\bf 25} (1984) 91.
%%CITATION = ZEPYA,C25,91;%%

%\cite{Post:2001si}\cite{Bijnens:2003uy}
\bibitem{Post:2001si}
P.~Post and K.~Schilcher,
%``K(l3) form factors at order p**6 in chiral perturbation theory,''
Eur.\ Phys.\ J.\ C {\bf 25} (2002) 427
[arXiv:hep-ph/0112352].
%%CITATION = HEP-PH 0112352;%%

%\cite{Bijnens:2003uy}
\bibitem{Bijnens:2003uy}
J.~Bijnens and P.~Talavera,
%``K(l3) decays in chiral perturbation theory,''
Nucl.\ Phys.\ B {\bf 669} (2003) 341
[arXiv:hep-ph/0303103]; see also 
http://www.thep.lu.se/$\sim$bijnens/chpt.html.
%%CITATION = HEP-PH 0303103;%%

%\cite{Bijnens:1998yu}
\bibitem{Bijnens:1998yu}
J.~Bijnens, G.~Colangelo and G.~Ecker,
%``Double chiral logs,''
Phys.\ Lett.\ B {\bf 441} (1998) 437
[arXiv:hep-ph/9808421].
%%CITATION = HEP-PH 9808421;%%

%\cite{Jamin:2004re}
\bibitem{Jamin:2004re}
M.~Jamin, J.~A.~Oller and A.~Pich,
%``Order p**6 chiral couplings from the scalar K pi form factor,''
JHEP {\bf 0402} (2004) 047
[arXiv:hep-ph/0401080].
%%CITATION = HEP-PH 0401080;%%

%\cite{Becirevic:2004ya}
\bibitem{Becirevic:2004ya}
D.~Becirevic et al.,
%``The K $\to$ pi vector form factor at zero momentum transfer on the
%lattice,''
Nucl.\ Phys.\ B {\bf 705} (2005) 339
[arXiv:hep-ph/0403217]; \\
%%CITATION = HEP-PH 0403217;%%
%\cite{Okamoto:2004df}
%\bibitem{Okamoto:2004df}
  M.~Okamoto  [Fermilab Lattice Collaboration],
  {\it Full CKM matrix with lattice QCD},
  arXiv:hep-lat/0412044.
  %%CITATION = HEP-LAT 0412044;%%


\bibitem{HNHonnef}
H. Neufeld, {\it Isospin violation in semileptonic decays},
Talk presented at the Workshop on Effective Field Theories
in Nuclear, Particle and Atomic Physics, Bad Honnef, Germany, Dec. 2004
and private communication.

\bibitem{Czarnecki:2004cw}
A.~Czarnecki, W.~J.~ Marciano and A.~Sirlin,
%"Precision measurements and CKM unitarity",%
Phys.\ Rev.\ D {\bf 70} (2004) 093006 [arXiv:hep-ph/0406324].
%%CITATION = HEP-PH 0406324;%%"

%\cite{Gamiz:2004ar}
\bibitem{Gamiz:2004ar}
  E.~Gamiz, M.~Jamin, A.~Pich, J.~Prades and F.~Schwab,
  %``V(us) and m(s) from hadronic tau decays,''
  Phys.\ Rev.\ Lett.\  {\bf 94} (2005) 011803
  [arXiv:hep-ph/0408044].
  %%CITATION = HEP-PH 0408044;%%

%\cite{Marciano:2004uf}
\bibitem{Marciano:2004uf}
W.~J.~Marciano,
%``Precise determination of $|$V(us)$|$ from lattice calculations of
%pseudoscalar decay constants,''
Phys.\ Rev.\ Lett.\  {\bf 93} (2004) 231803
[arXiv:hep-ph/0402299].
%%CITATION = HEP-PH 0402299;%%

%\cite{Callan:1966hu}
\bibitem{Callan:1966hu}
  C.~G.~Callan and S.~B.~Treiman,
  %``Equal Time Commutators And K Meson Decays,''
  Phys.\ Rev.\ Lett.\  {\bf 16} (1966) 153; \\
  %%CITATION = PRLTA,16,153;%%
  R.~F.~Dashen and M.~Weinstein,
  %``Theorem On The Form-Factors In K-L-3 Decay,''
  Phys.\ Rev.\ Lett.\  {\bf 22} (1969) 1337.
  %%CITATION = PRLTA,22,1337;%%

\bibitem{Serebrov:2004zf}
A.~Serebrov et al.,
%"Measurement of the neutron lifetime using a gravitational
%trap and a low-temperature Fomblin coating",
Phys.\ Lett.\ B {\bf 605} (2005) 72 [arXiv:nucl-ex/0408009].
%%CITATION = NUCL-EX 0408009;%%"
%%%%%%%%%%%%%%%%%%%%%%%%%%%%%%%%%%%%%%  
% %% %% %%

\end{thebibliography}
\end{document}